\documentclass[prb,singlecolumn,a4paper,amsmath,amssymb]{revtex4}

\usepackage{amsfonts}
\usepackage{graphicx}
\usepackage[usenames]{color}
\usepackage[utf8]{inputenc}
\usepackage{hyperref}
\usepackage{caption}
\usepackage{array}
\usepackage{pdfpages}

\newcommand{\be}{\begin{equation}}
\newcommand{\ee}{\end{equation}}

\newcommand{\kb}{k_\mathrm{_B}}

\begin{document}

\title{Problem of the 8-th Experimental Physics Olympiad, Skopje, 8 May 2021
Determination of Planck constant by LED}

%%%%%%%%%%%%%%%%%%%%%%%%%%%%%%%%%%

\author{Todor~M.~Mishonov, Aleksander~P.~Petkov, Matteo Andreoni, Emil~G.~Petkov, Albert~M.~Varonov}
\email[E-mail: ]{epo@bgphysics.eu}
%\affiliation{Physics Faculty, St.~Clement of Ohrid University at Sofia, 
%5 James Bourchier Blvd., BG-1164 Sofia}
\affiliation{Georgi Nadjakov Institute of Solid State Physics, Bulgarian Academy of Sciences\\
72 Tzarigradsko Chaussee Blvd., BG-1784 Sofia, Bulgaria}
\author{Iglika~M.~Dimitrova}
\affiliation{Faculty of Chemical Technologies, Department of Physical Chemistry,
University of Chemical Technology and Metallurgy,
8 Kliment Ohridski Blvd., BG-1756 Sofia}
\affiliation{Georgi Nadjakov Institute of Solid State Physics, Bulgarian Academy of Sciences\\
72 Tzarigradsko Chaussee Blvd., BG-1784 Sofia, Bulgaria}%
\author{Leonora Velkoska}
\author{Riste Popeski-Dimovski}
\email{ristepd@gmail.com}
\affiliation{Institute of Physics, Faculty of Natural Sciences and Mathematics, 
``Ss. Cyril and Methodius'' University, Skopje, R.~N.~Macedonia}
%\author{Aleksander~P.~Petkov, Matteo Andreoni, Emil~G.~Petkov, Albert~M.~Varonov}
%\affiliation{Georgi Nadjakov Institute of Solid State Physics, Bulgarian Academy of Sciences\\
%72 Tzarigradsko Chaussee Blvd., BG-1784 Sofia, Bulgaria}%

\date{May 31, 19:59}

\begin{abstract}
This is the problem of the 8$^\mathrm{th}$ International Experimental Physics Olympiad (EPO).
The task of the EPO8 is to determine Plank constant $\hbar=h/2\pi$
using the given set-up with LED.
If you have an idea how to do it, do it and send us the result;
skip the reading of the detailed step by step instructions 
with increasing difficulties.
We expect the participants to follow the 
suggested items -- they are instructive for physics education in general.
Only the reading of historical remarks given in the first section can be 
omitted during the Olympiad without loss of generality.
Participants should try solving as much tasks as they can
without paying attention to the age categories: give your best.
\end{abstract}
\captionsetup{labelfont={normalsize},textfont={small},justification=centerlast}
\maketitle
%%%%%%%%%%%%

\section{EPO -- a historical perspective. Skip this section during the Olympiad}

From its very beginning, the Experimental Physics Olympiad (EPO) is worldwide known;
all Olympiad problems have been published in Internet~\cite{EPO1,EPO2,EPO3,EPO4,EPO5,EPO6,EPO7} 
and from the very beginning there were 120 participants.
In the last years high-school students from more than 10 countries participated and the distance between the most distant cities 
is more than 4~Mm.

Let us describe the main differences between EPO and other similar competitions.
\begin{itemize}
\item Each participant in EPO receives as a gift from the organizers the set-up, 
which one worked with.
In such a way, after the Olympiad has finished, even bad performed participant is able to repeat the experiment and reach the level of the champion.
In this way, the Olympiad directly affects the teaching level in the whole world.
After the end of the school year, the set-up remains in the school, where the participant has studied.
\item Each of the problems is original and is connected to fundamental physics or the  understanding of the operation of a technical patent.
\item The Olympic idea is realized in EPO in its initial from 
and everyone willing to participate from around the world can do that.
There is no limit in the participants number.
On the other hand, the similarity with other Olympiads is that the problems are direct illustration of the study material and alongside with other similar competitions mitigates the secondary education degradation, which is a world tendency.
\item One and the same experimental set-up is given to all participants but the tasks are different for the different age groups, the same as the swimming pool water is equally wet for all age groups in a swimming competition.
\item One of the most important goals of the Olympiad is the student to repeat the experiment at home and to analyze the theory necessary for the understanding.
In this way any even badly performed motivated participant has the possibility to be introduced to the corresponding physics field, even though there is no physics classroom in his/her school, even though the physics education in his/her country to be deliberately destroyed.
\end{itemize}

We will briefly mention the problems of former 7 EPOs: 
1) The setup of EPO1~\cite{EPO1} was actually a student version of the American patent for auto-zero
and chopper stabilized direct current amplifiers.
It was notable that many students were able to understand the operation of an American patent without special preparation.\cite{chopper}
2) The problem of EPO2~\cite{EPO2} was to measure Planck constant by diffraction of a LED light by a compact disk.
3) A contemporary realization of the assigned to NASA patent for the use of negative impedance converter for generation of voltage oscillations was the set-up of EPO3.~\cite{EPO3}
4) EPO4~\cite{EPO4} was devoted to the fundamental physics -- to determine the speed of light by measuring electric and magnetic forces.
The innovative element was the application of the catastrophe theory in the analysis of the stability of a pendulum.
5) The topic of the EPO5~\cite{EPO5} was to measure the Boltzmann constant $\kb$ following the Einstein idea of study thermal fluctuations of electric voltage of a capacitor.
6) The EPO6~\cite{EPO6} problem can be considered as a continuation of the previous Olympiad.
With a similar electronic circuit Schottky noise is measured and his idea for the determination of the electron charge is realized.
7) The EPO7~\cite{EPO7} problem was to measure a large inductance made by a general impedance converter by the Maxwell-Wien bridge.
Each problem given at EPO can be considered as a dissertation in methodology of physics education.

In short, the established traditions is a balance between contemporary working technical inventions and fundamental physics.

The EPO problems are meant for high school and university students 
but are posed by teachers with co authorship 
with colleagues working in universities or scientific institutes.
For colleagues interested in new author's problems for the needs of the contemporary physics education we share our experience in the description of the experimental set-ups described at a university level.
These are for instance:

1) The determination of the Planck constant without light but only with electronic processes study;~\cite{EJP_Planck} this set-up requires the usage of an oscilloscope but in some countries the oscilloscopes are available in  in the high schools physics labs and the prices of the former is constantly going down.

2) The speed of the light without the usage of scales or high frequency equipment is another innovative set-up~\cite{EJP_light} for high school education.
And the idea for this experiment is given by our teacher in electrodynamics Maxwell.

3) In the physics curriculum in all countries it is mentioned that the temperature is a measure of average kinetic energy of the gas molecules but the Boltzmann constant $k_\mathrm{B}$ that gives the relation between energy and temperature is not measured in high school and even rarely in the best universities.
The experimental set-up for $k_\mathrm{B}$ by the method proposed by Einstein (the EPO5 problem)
is described~\cite{EJP_Boltzmann} as a set-up for university school lab exercise in a impact factor journal.
But what larger impact an experimental set-up that is used by high school students from Kazakhstan and Macedonia and the surrounding countries can have.
More than 100 set-ups were distributed around the world.

4) Similar thoughts can be expressed for the electron charge $q_e$.
This fundamental constant is also mentioned in the high school education as a humanitarian incantation but is not measured.
We broke this tradition and described an experimental set-up (from EPO6) in the European Journal of Physics.~\cite{EJP_Schottky}
This set-up can be built for a week in every high school.
The Schottky idea for determination of the electron charge by measuring voltage fluctuations is used.
From the idea to the realization more than 100 years have passed and one of the reasons is that in many countries the largest enemy of the education is the ministry of education.

The development of some set-ups required additional study of circuits with operational amplifiers (OpAmps).
This led to introduction of the master equation of OpAms
applied to the stability of circuits with Negative Impedance Converter
(NIC),~\cite{NIC} 
Generalized Impedance Converter~\cite{RSI_GIC}
and study of Probability Distribution Function~\cite{PDF}
of the crossover frequency of OpAmps.

The mission of the physics teachers in the worldwide progress is evident -- precisely our science reshaped the world in the last century.
Successful innovative EPO set-ups after some update can be manufactured by companies specialized in production of educational equipment like 
TeachSpin~\cite{TeachSpin} and PHYWE~\cite{PhyWe} for instance. 

\section{Description of the experimental set-up 
and the conditions for online participants}
\begin{figure}[ht]
\includegraphics[scale=0.32]{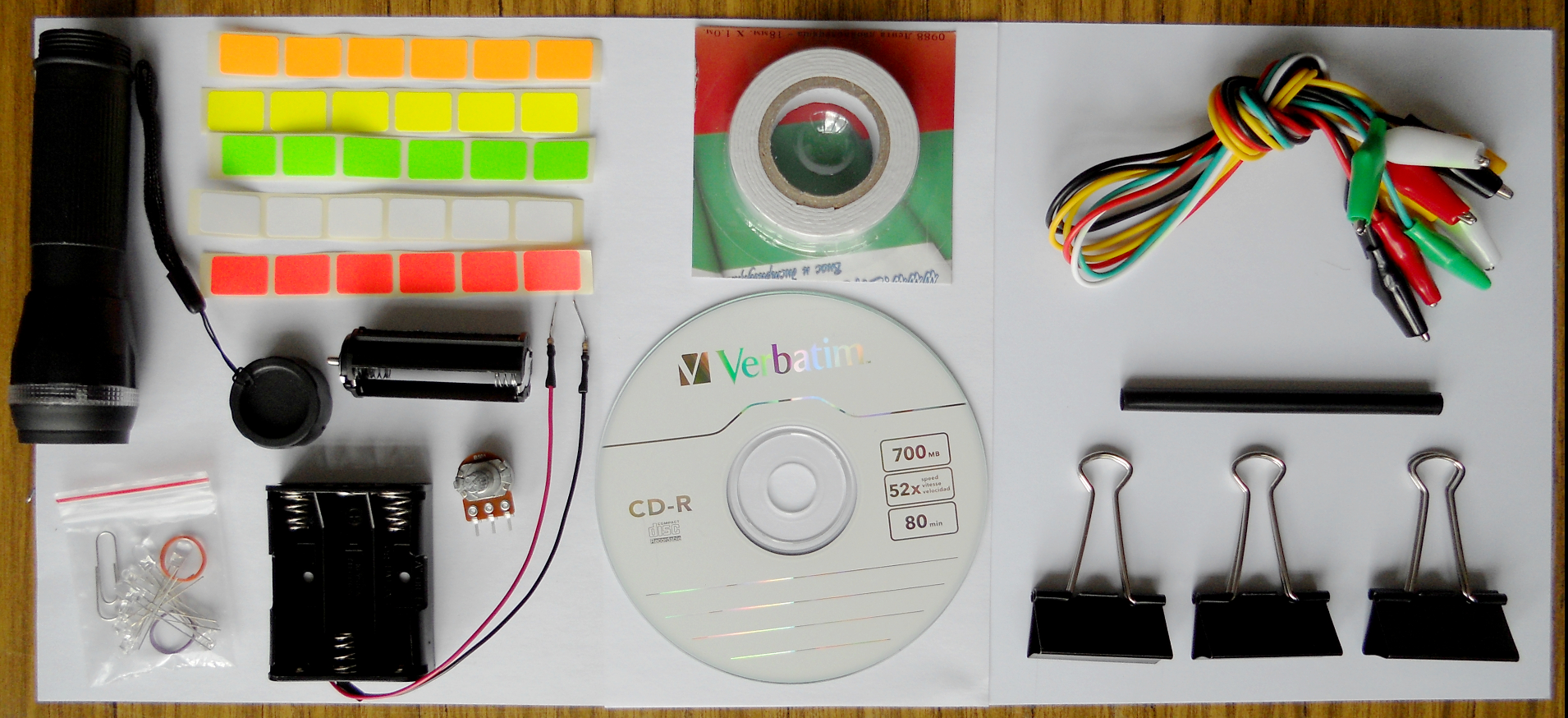}
\caption{Details of the set-up given to the participants:
1 small plastic bag containing 8 light emitting diodes (LED),
1 paper clip and 2 small rubber rings,
1 potentiometer, 1 piece of black plastic straw,
3 metal binder clips,
1 battery holder for 3~AA-type batteries,
1 double sided adhesive tape,
1 compact disk (CD),
7 connecting crocodile cables,
1 torch with battery holder taken out,
5 bands of 6 differently colored sticky labels.}
\label{fig:set-up}
\end{figure}

\begin{figure}[ht]
\includegraphics[scale=0.48]{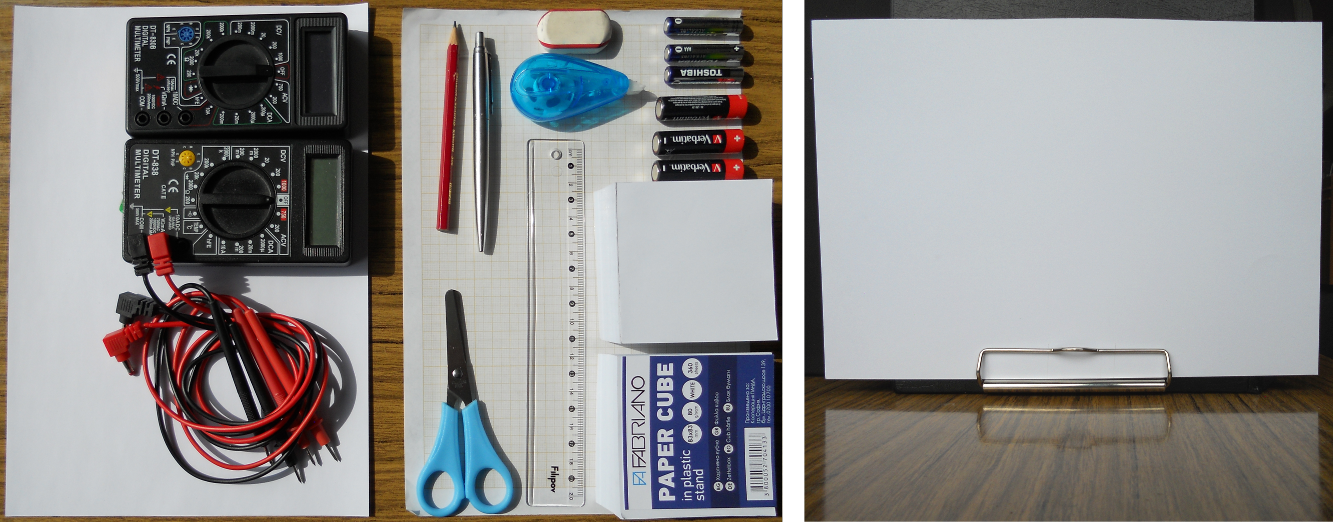}
\caption{
Details which every participant have to ensure himself:
computer with Internet connection,
mobile phone, a camera or a scanner to create a PDF or picture file
of your paper work during the Olympiad.
Additionally:
1) 2 multimeters with original connecting cables,
2) 3 AA type batteries of 1.5 V,
3) 3 AAA type batteries of 1.5 V,
4) pen(s) and a plastic ruler,
5) 2 (or 1 if height of cube $>$6~cm) paper cubes 8$\times$8~cm 
6) scissors,
7) at least 2 sheets of millimeter paper or squared paper sheets
and 4 white sheets of ordinary A4 paper,
8) 2 sheets of A4 white paper with 240~g/m$^2$ or larger thickness,
9) calculator (not shown here and highly recommended :-) ,
10) pad for A4 paper (recommended),
11) pencil, rubber eraser, white band eraser (recommended),
*) sandwiches and bottle of water (strongly recommended).
0) Nothing excessive on the cleaned working table and the room has to be darkened.
}
\label{fig:on_spot_details}
\end{figure}

The set-up you received is represented in the Fig.~\ref{fig:set-up}.
Ones again we repeat the condition that 
the room in which you work can be darken by dense curtains, for example.
%%%%%%%%%%%%%%
%%%%%%%%%%%%%%
The details which every participant has to ensure 
are depicted in Fig.~\ref{fig:on_spot_details}.
%%%%%%%%%%%%%%%%
%%%%%%%%%%%%%%%%
In the next sections different tasks with increasing difficulty 
are described in the sections corresponding to every age category.
The jury will looks at the experimental data, tables and graphics.
It is not necessary to write the humanitarian text between them,
only mark the number of the corresponding task.
We wish you success.

%%%%%%%%%%%%%%%%
\section{Tasks S. Getting to know the voltage source}
\label{Sec:beginning}

\begin{enumerate}

\item 
\label{r+-}
Turn on the multimeter as an Ohm-meter ($\Omega$).
With maximum accuracy measure the 2 soldered resistors to the triple AA battery holder and write down the results.
What is the resistance $r_+$ and $r_-$ of the resistors at the (+) 
and (-) ends of the 3 battery holder?
Calculate $r=r_++r_-.$

\item 
\label{R_LR}
What is the resistance $R_\mathrm{LR}$ 
between the left and right ends of the potentiometer?

\item
\label{R_pot}
Measure the resistance of the potentiometer between the middle terminal and one of the others.
In what interval does the resistance change
when you rotate its axis $(R_\mathrm{min}, R_\mathrm{max})$?

\item 
\label{E123}
Turn on the multimeter as a DCV and measure the voltages of your 1.5~V AA batteries.
Order them if they are different
$\mathcal{E}_1\le\mathcal{E}_2\le\mathcal{E}_3.$

\item 
\label{Etot}
Put the batteries at battery holder and measure 
the total electromotive force $\mathcal{E}$.
What is the accuracy of the relation 
$\mathcal{E}=\mathcal{E}_1+\mathcal{E}_2+\mathcal{E}_3?$

\item 
\label{167mV}
Calculate how many mV is the difference between $\frac12\,$V and $\frac13\,\mathrm{V}$?

\item 
\label{Umax}
Using crocodile cables, battery holder, potentiometer and voltmeter 
connect the voltage source circuit depicted in Fig.~\ref{fig:V_source}.
%%%%%%%%%%%%%%
\begin{figure}[ht]
\includegraphics[scale=0.4]{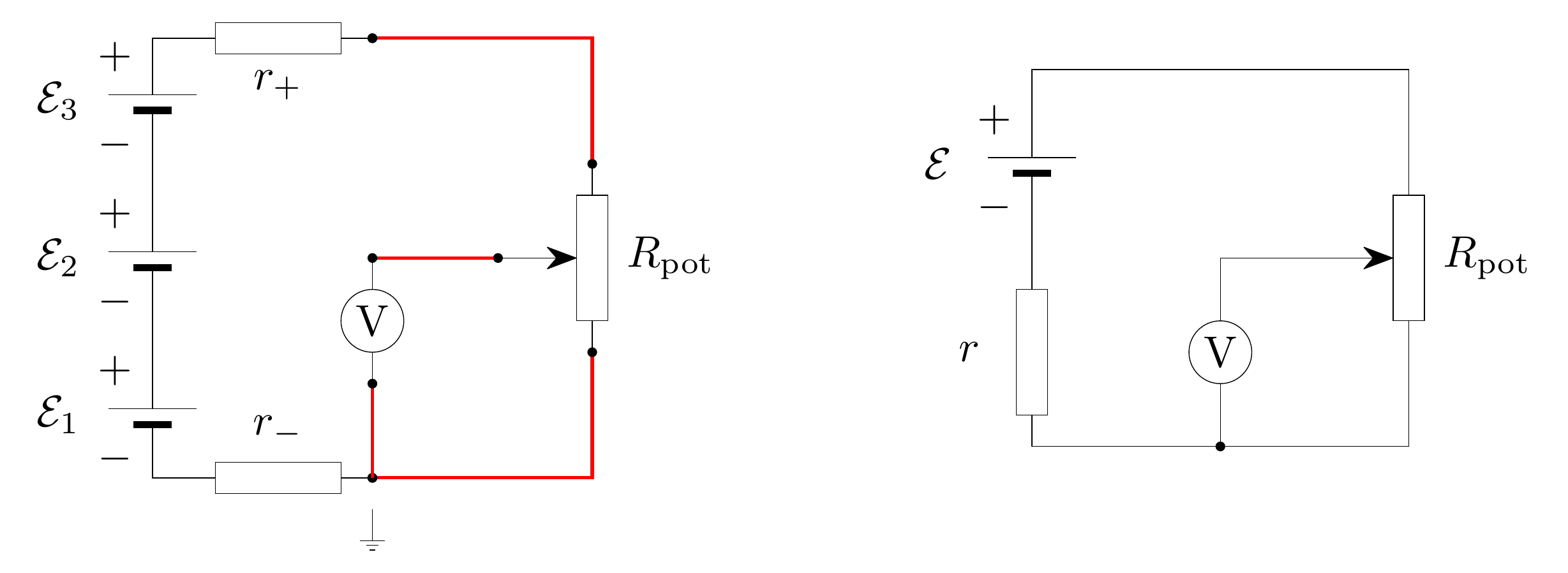}
\caption{Left. Detailed circuit with: battery holder with 3 AA type batteries,
two resistors, potentiometer, and voltmeter. 
Connecting 4 cables are given by red color
and crocodiles clips are marked by points (dots).
Right. Schematic representation of the same circuit:
voltage source with electromotive voltage 
$\mathcal{E}=\mathcal{E}_1+\mathcal{E}_2+\mathcal{E}_3$
and sequential internal resistor $r=r_-+r_+$
corresponding to the battery holder with the 3 AA type batteries inserted in,
potentiometer with resistance $R_\mathrm{pot}$,
and voltmeter.
}
\label{fig:V_source} 
\end{figure}
%%%%%%%%%%%%%%
Determine the the interval in which the voltage
measured by the voltmeter changes
$U_\mathrm{V}\in(0,U_\mathrm{max})$
when you rotate the axis of the potentiometer
(ensure $U_\mathrm{V}>0$).
You can use a piece of double side adhesive band
to fix the potentiometer on the pad.
%With additional piece of adhesive band
%using only one side you can make a handle to the
%potentiometer.
%If necessary exchange the cables to the voltmeter
%in order positive end of the batteries to correspond
%to positive voltage.

\item
Using the circuit from Fig.~\ref{fig:V_source},
connect two additional crocodile cables to the 
middle terminal of the potentiometer
and the ground point of the circuit, leaving the other connectors of the two cables open.
In such a way the voltmeter shows 
the voltage applied to the LED $U=U_\mathrm{V}$.

\item
Note that light-emitting diodes (LED) have terminals with different lengths;
\textit{the longer (anode) has to be connected to positive voltage}.
Connect the open ends of the two cables to one of the LEDs,
following Fig.~\ref{fig:U*}.
%%%%%%%%%%%%%%
\begin{figure}[ht]
\includegraphics[scale=0.5]{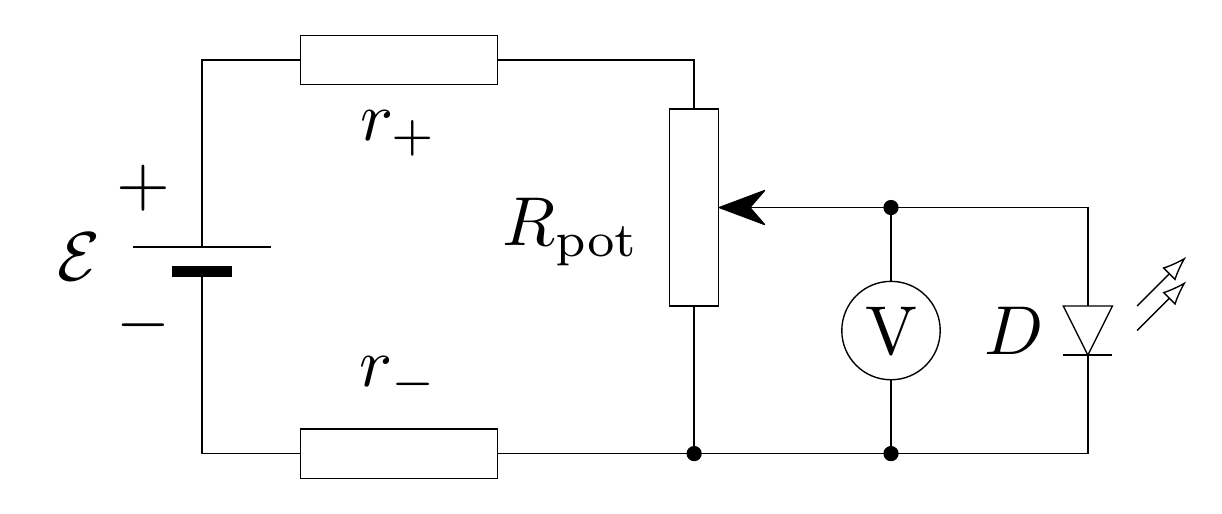}
\caption{LED connected to the voltage source 
from Fig.~\ref{fig:V_source},
circuit for measuring the voltage $U^*$ at which the LED emits the weakest evanescent light.
}
\label{fig:U*} 
\end{figure}
%%%%%%%%%%%%%%
Do you see the similarities and difference between in the connections in Fig.~\ref{fig:V_source} and Fig.~\ref{fig:U*}?

\item
\label{enum}
Turn on the voltage source until the connected LED begins emitting clear visible light.
Make a table with 3 columns,
column 1 is the sequential number of the LED,
column 2 is its color,
column 3 is $U^*$ the smallest voltage at which the LED emits light.
From now on you have to keep the enumeration and order of the LEDs,
you can use the colored sticky labels for that 
but make sure you can easily remove the labels if necessary.
%written on the square cardboard and the second is the color.

\item
\label{item:U*}
Now it is good to darken the room.
Put the connected LED into the one end of the plastic straw.
Measure $U^*$ by carefully rotating the handle of the potentiometer
and visually observing with your eye 
through the other end of the plastic straw
when the LED begins emitting light
(the plastic straw is a waveguide for the LED light).
We recommend at least 5 different measurements and taking the median (or middle) value for $U^*$.
In this way determine $U^*$ for all available and working LEDs and fill column 3 of the table.
The human eye is a very good sensor however the results are subjective.

%%give maximal voltage.
%%There will be some diodes with slightly different colors.
%%Be careful, you can easily burn all diodes.
%To the diodes you have look trough the black tube.
%The purpose of this task is to measure 
%the lighting voltage of the diodes $U^*$.
%%In other words $U^*$ is the smallest voltage at which
%%the LED emits the light
%%or bigger voltage at which LED is dark.
%The eye is a very good sensor however 
%the results are subjective.
%Carefully rotate the handle of the potentiometer and for every LED
%make at least 5 different measurements.
%Determine $U^*$ as median of the different measurements.
%For example voltage which is bigger than 2 on the smallest measurements
%and smaller than 2 biggest one.
%The third column of your table has to be lighting voltage $U^*$,
%i.e. number, color, voltage.
%The corresponding circuit is represented at Fig.~\ref{fig:U*}

%%%%%%%%%%%%%%%%%%%%
\section{Tasks M. Diffraction of LED light}
%%%%%%%%%%%%%%%%%%%%

The purpose of this set of tasks is to measure the the wavelength $\lambda$
and the frequency of the light $\nu=c/\lambda$ of every LED;
where $c\approx 300\,\mathrm{Mm/s}$ is the speed of light.
For diffraction grating we will use Compact Disk (CD) with constant
$d=1600\,\mathrm{nm}$.

\item
\label{square}
Cut a 8$\times$8~cm square piece of cardboard or thicker paper.
Using the double adhesive tape carefully stick all measured LEDs in \ref{item:U*}
on this square piece as shown in Fig.~\ref{fig:square8diodes}
%%%%%%%%%%%%%%
\begin{figure}[h]
\includegraphics[scale=0.4]{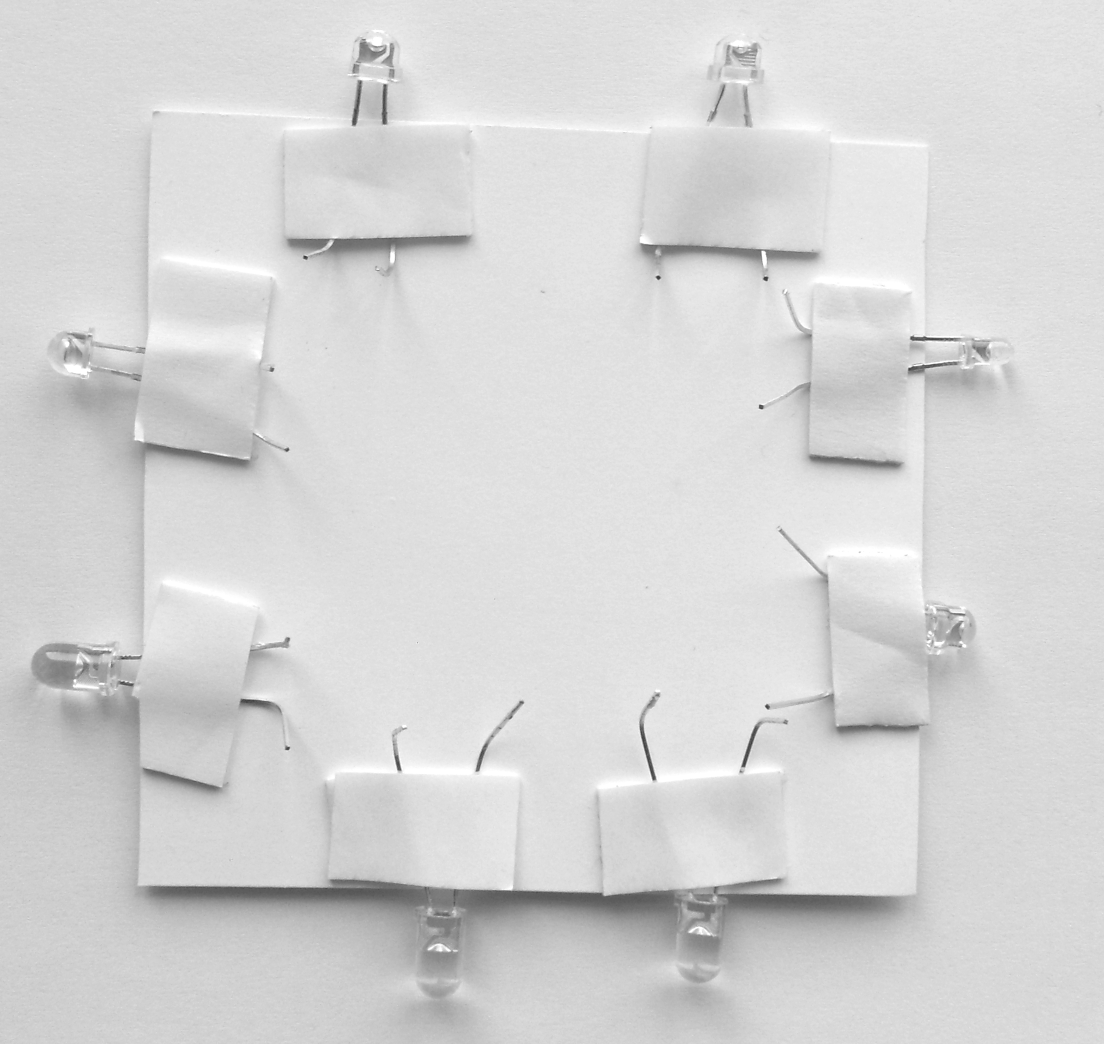}
\caption{Square piece with 8 LED stuck by double sided adhesive tape.
Enumerate each LED and write a
small (+) sign close to the number to mark the anode of LED.
The terminals are bent in vertical direction in order to be bitten by crocodiles :-).
%Look at the LED trough the black straw in order to observe the minimal light LED and to measure the corresponding voltage $U^*$.
}
\label{fig:square8diodes} 
\end{figure}
%%%%%%%%%%%%%%
and enumerate them according to \ref{enum}.
It is better the ends of the diodes to be bent vertically in order to be connected 
by crocodile cables easier.
For further convenience, try using one and the same orientation of the terminals,
or mark with (+) on the cardboard the longer terminal.
If you cannot accomplish this item, you can still solve the whole problem,
therefore proceed to the next tasks.

\item
Now you need a diffraction grating. 
In order to obtain it you have to remove the metallic foil layer from the CD
as it is described in Fig.~\ref{Fig:set_disk_foil_detachment}.
%%%%%%%%%%%%%%
\begin{figure}[ht]
\includegraphics[scale=0.3]{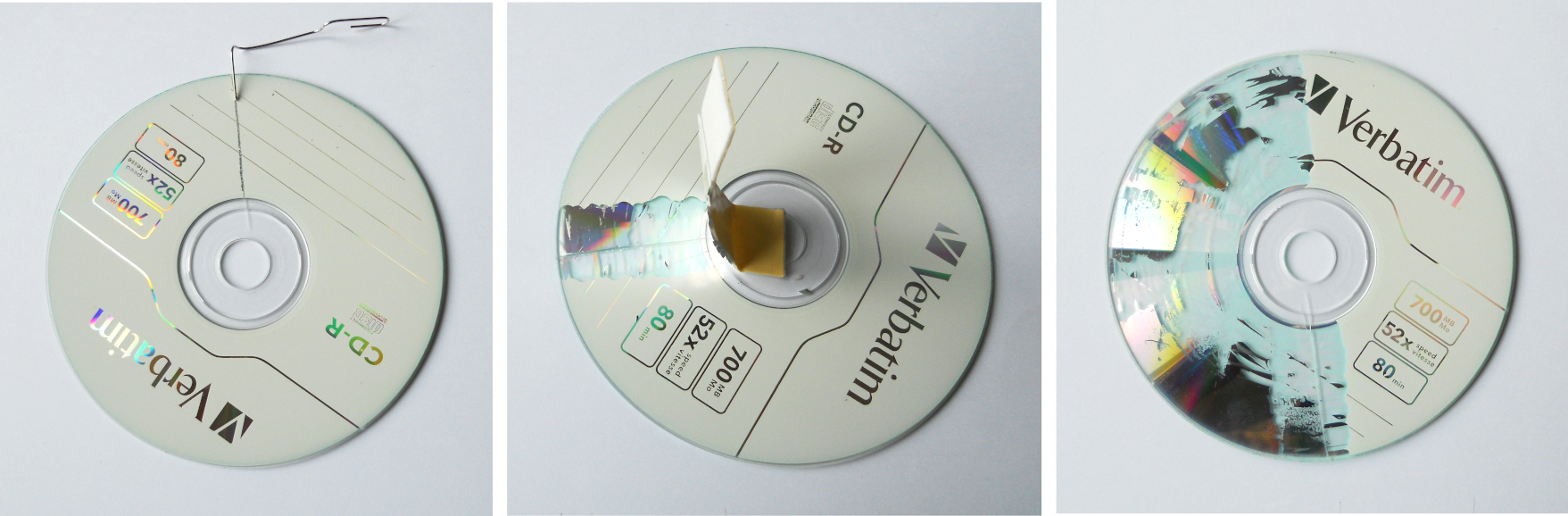}
\caption{The technology of peeling the foil of the compact disk:
1) The foil is scratched radially with the paper clip.
2) A piece of the adhesive tape is stuck on the scratch.
3) Then the piece is pulled and the foil stuck on the tape is removed from the CD.
4) Repeat steps 2) and 3) above (or near) the edges of the remaining foil 
until about half part of the disc is peeled from the foil
as a half Moon at the right.
You can use both sides of the double adhesive band.}
\label{Fig:set_disk_foil_detachment} 
\end{figure}

\item 
Put the 3 AAA type batteries in the torch battery holder,
assemble the torch, turn it on and focus it at infinity
(make the beam as narrow as possible).
You have almost parallel beam of white light.
Using the different halves of the disk, you can observe a spectrum 
of the torch white LED at reflection and transmission.
In order to observe spectrum at transmission,
the light from the torch has to pass through the peeled half of the disc
and fall on a blank sheet of paper.
%the CD and a screen from a cardboard and white sheet of paper
%or thicker white sheet of paper have to be fixed by a paper binder clip 
%as it is explained in the schemes in Fig.~\ref{Fig:setup_scheme} 
%and photograph Fig.~\ref{Fig:experimental_setup_blue_amber}.
The sheet of paper and the CD have to be parallel and we recommend the distance 
between them $D$ to be approximately 15~cm.
Try obtaining central maximum and two horizontal spectra with the torch; 
this is the initial exercise to the more difficult task with the set of LEDs.

\item
Take out the lens from the torch as shown in Fig.~\ref{Fig:set_diode_lens_disk} .
%%%%%%%%%%%%%%
\begin{figure}[ht]
\includegraphics[scale=0.3]{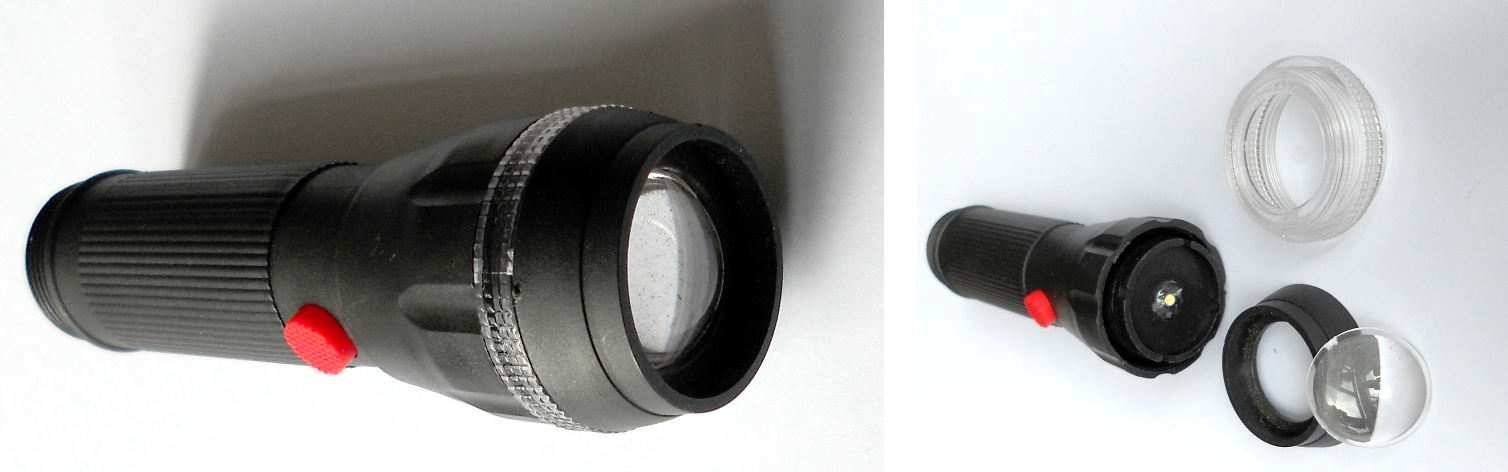}
\caption{Torch with lens: carefully unscrew the part holding the lens to the torch, disassemble it and take out the lens.}
\label{Fig:set_diode_lens_disk} 
\end{figure}
%%%%%%%%%%%%%%

\item
Fix the lens at the top of 1 metal binder clip with rubber rings
as it is shown in Fig.~\ref{Fig:setup_scheme} and 
Fig.~\ref{Fig:experimental_setup_blue_amber}.
Stick (highly recommended) the binder clip with the lens on the pad
(or table) with a piece of double sided adhesive tape.

\item
Use the paper cubes (or cube) and adjust the height approximately equal
to the height of the lens on the binder clip.

\item
Place the 8$\times$8~cm squared piece with LEDs made in \ref{square} on the cubes,
bring them together near the lens, so that one of the LEDs to face the lens and be at close distance behind it
as shown in both
Fig.~\ref{Fig:setup_scheme} and 
Fig.~\ref{Fig:experimental_setup_blue_amber}.
You can use books instead of the paper cubes but height change is more difficult.
%%%%%%%%%%%%%%
\begin{figure}[h]
\includegraphics[scale=0.32]{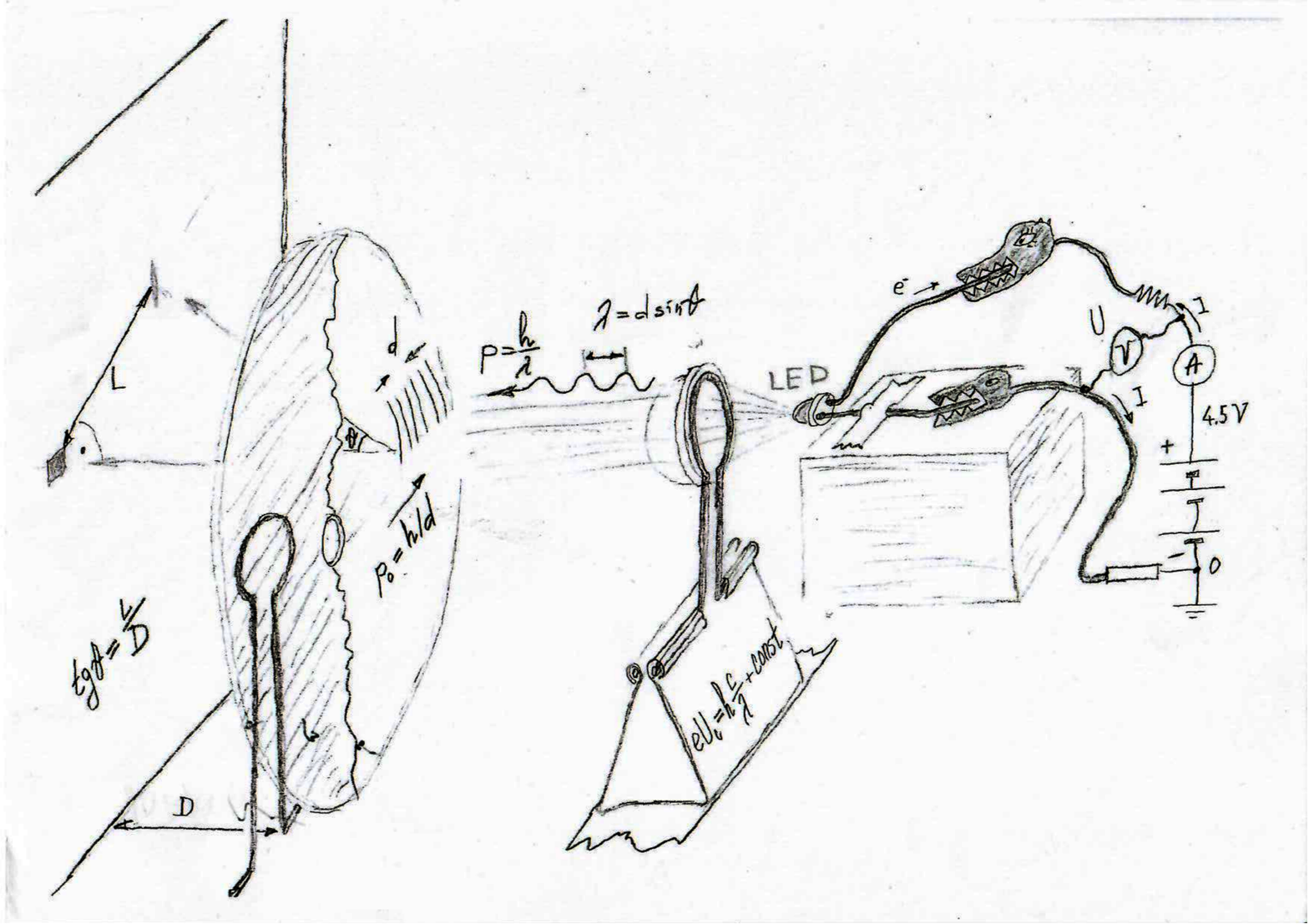}
\caption{Scheme of the setup shown on 
Fig.~\ref{Fig:experimental_setup_blue_amber}.
The LED is connected with wires to the voltage supply source using crocodile clips.
A photon from the collimated light has momentum $p=h/\lambda$, which is perpendicular to the CD.
The photon obtains momentum $p_0=h/d$ in direction perpendicular to the direction of the tracks of the CD, 
and for the diffraction angle $\theta$ we have $\tan \theta=p_0/p$.
$D$ is the distance between the screen and the CD, and $L$ is the distance between the central and the first diffraction maximum; $\tan\theta=L/D.$ 
The binder clips, which hold the lens and the CD, 
are drawn schematically, too.}
\label{Fig:setup_scheme} 
\end{figure}
%%%%%%%%%%%%%%
%%%%%%%%%%%%%%
\begin{figure}[h]
\includegraphics[scale=0.2]{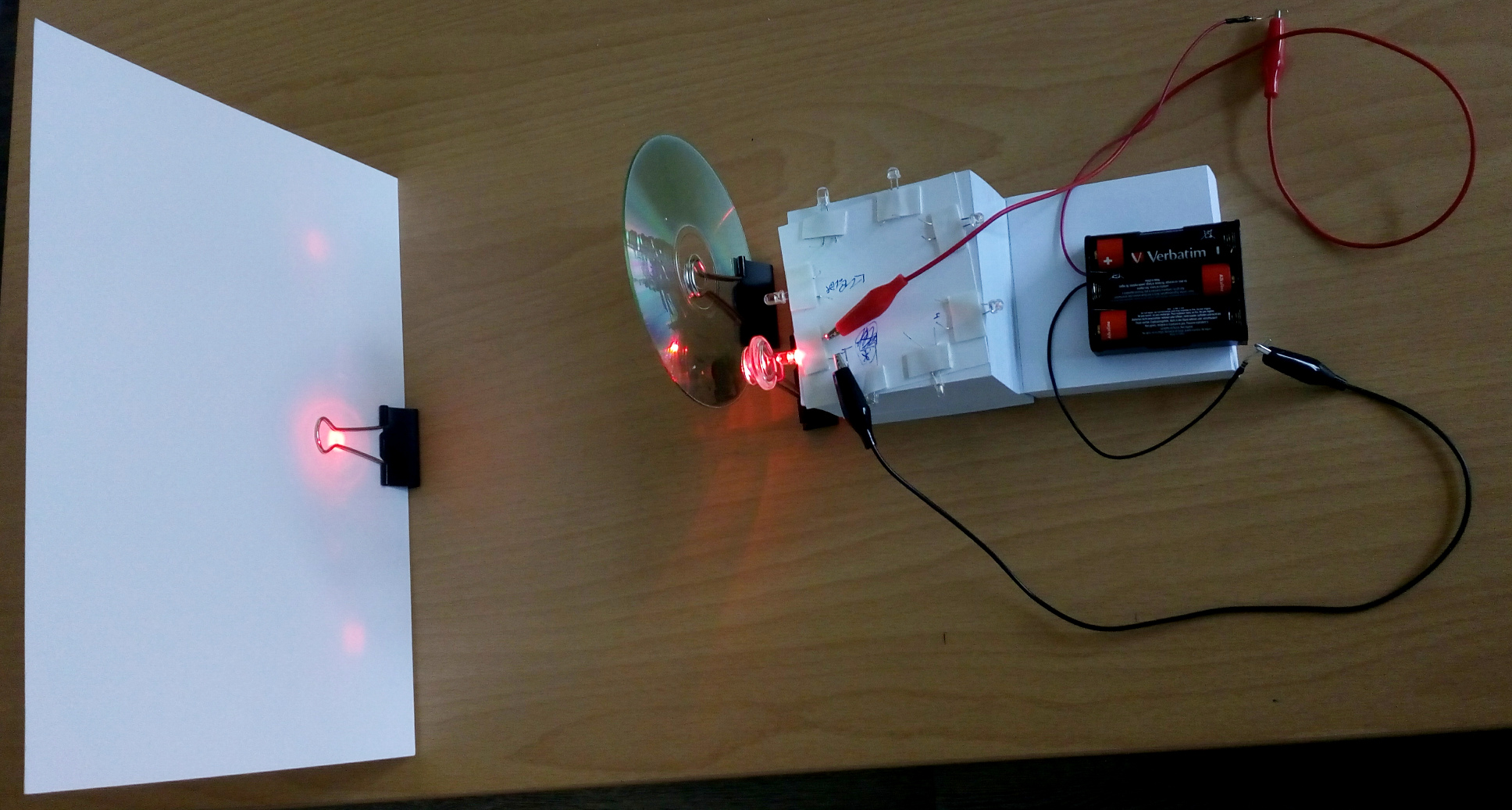}
\caption{Experimental set-up for measuring the wavelength of LED light.
The red LED is connected with the connecting cables to the voltage supply source.
The light is collimated with the lens then it passes through the tracks 
of the transparent compact disk
and one can see a bright image of the LED on the screen.
Together with the central maximum two pale diffraction maximums can be observed on the screen.
Diffraction angles are different for the different colors from the set. 
The wavelength $\lambda$ of the light 
can be determined by measuring the diffraction angles $\theta$.
}
\label{Fig:experimental_setup_blue_amber} 
\end{figure}
%%%%%%%%%%%%%%

\item
\label{diffr}
Put the CD in the second metal binder clip and place it after the lens
as shown in both 
Fig.~\ref{Fig:setup_scheme} and 
Fig.~\ref{Fig:experimental_setup_blue_amber}.
The binder clip with the lens must be between the diodes and the binder clip with the CD.

\item
\label{screen}
Put the cardboard box with a normal sheet of paper or
the thicker sheet of paper in the third metal binder clip.
This is your screen.

\item
Carefully arrange the diodes, binder clip with lens, binder clips with CD and screen
as shown in both
Fig.~\ref{Fig:setup_scheme} and 
Fig.~\ref{Fig:experimental_setup_blue_amber}.
Screen, CD and lens should be parallel to each other and
perpendicular to the direction of propagation of light
It is best to use the periphery of the CD for the diffraction region. 

\item
Use the voltage source and apply $\mathcal{E}$ to one of the red LED 
and place it behind the lens.
Make additional adjustments in the arrangement
until you see the bright central maximum and pale diffraction maximums
on your screen as in Fig.~\ref{Fig:experimental_setup_blue_amber}.
The focus distance is very small and putting the LED close to the lens
you can see the real image of the lighting part of the led on the screen.
The method for adjusting of the set-up is individual, we recommend
all details to be on the paper pad where with pen you can mark
the place of every detail which can be easily returned in the same place
after removal.

\item
Try to make fine tuning of the set-up slightly rotating or moving the screen and/or CD.
Additionally adjust the height of the square piece with LEDs if necessary. 
The distances between central maximum and left $L_\mathrm{l}$ 
and right one $L_\mathrm{r}$ should
be approximately equal and approximately on a horizontal line.
Experimentally the mean value $L=\frac12(L_\mathrm{l}+L_\mathrm{r})$
can be determined as half of the distance between the diffraction maximums.

%At the same height should be the center of the disk;
%change its position in the clips.
%Put the disk on the path off the light and you will see
%the bright central maximum and pale diffraction maximums.
%For the diffraction region use the periphery of the disk. 

\item
For all of LEDs you have to measure $L$ and 
the distance $D$ between the screen and CD,
which is better to be kept constant for each LED with the adhesive tape.
Then we can calculate sine of the diffraction angle $\theta=\arctan(L/D)$
using the distances
$\sin\theta=L/\sqrt{L^2+D^2}=\sin(\arctan(L/D))$.
The wavelength then is given by the well-known formula
\be
\lambda=d\sin\theta.
\ee
Now you can calculate the frequency $\nu=c/\lambda=1/T$ and the period $T$ of the oscillations.

\item
The determination of the frequency of light of all LEDs
is the main goal of the optical part of the task.
We recommend the measured and calculated 
data for all diodes to be represented in Table~\ref{tbl:LED}.
%1) number,
%2) color,
%3) $L$~[mm],
%4) $D$~[mm]
%5) $\theta$~[deg (degrees)],
%6) $\sin\theta$,
%7) $\lambda$~[nm],
%8) $\nu$~[THz],
%where: 1~mm=10$^{-3}\,$m, 
%1~nm=10$^{-9}\,$m, and
%1~THz$=10^{12}$~Hz.
%%%%%%%
\begin{table}[h]
\begin{tabular}{ c  c c  c c  c  c c c}
		\tableline \tableline
		&  \\ [-1em]
		$\#$  \hspace{1.5pt} & color & \hspace{5pt} $D$~[mm] & \hspace{1.5pt} $L$~[mm] & 							    \hspace{1.5pt} $\sin(\theta)$ & \hspace{1.5pt} $\theta$~[deg] & \hspace{1.5pt} $\lambda$~[nm] & \hspace{1.5pt} $\nu$~[THz] & \hspace{1.5pt} $q_eU^*$~[$10^{-19}$~J]
		 \\ \tableline 
			&  \\ [-1em] 
			1	&	\dots	&	\dots	&	\dots 	&	\dots &	\dots	&	\dots 	&	\dots &	\dots \\
			\dots	&	\dots	&	\dots	&	\dots	&	\dots &	\dots	&	\dots 	&	\dots &	\dots\\
\tableline \tableline
\end{tabular}	\caption{Table for arranging the measured results from the LED diffraction patterns.
We recommend not to draw the vertical lines of the table since they quite often make it unusable.}
	\label{tbl:LED}
\end{table}
%%%%%%%

\item
Now you can address the main goal of the Olympiad 
-- the determination of Planck constant $h$.
Using the electron charge $q_e=1.60\times 10^{-19}$~C 
complete the last column of Table~\ref{tbl:LED} 
by calculating the energies $q_eU^*$ for all LEDs, using the results from item~\ref{item:U*}.
Those energies are related to photon energies by the equation
\be
h\nu=q_eU^*+A.
\ee
Our estimation for Planck constant $h$ is based on the approximation that the
parameter $A$ is weakly dependent of the chemical compound of the LED semiconductor
therefore is almost constant.

\item
%Now you have a graphical task. 
Consider carefully the range of variables, the used units and their powers.
Mark in a graphic plot of millimeter paper:
($x$) in the abscissa the frequencies of the LED and
($y$) along the ordinate axis the energies $q_eU^*$.
You can use, for example, THz and $10^{-19}$~J.
In this $(\nu, q_eU^*)$ plot 
place the points corresponding to all measured LEDs.
For each point write down the number according to the table in item~\ref{item:U*}.

%\item The results from the measurement according to the instructions are written in the 3-rd column of Table~\ref{tbl:LED}.

\item 
\label{linear}
Perform linear regression of these points
(draw a straight line which passes closest to all points according to you).
It is allowed to ignore a point if it deviates too much from the others
and if you do it, write it down explicitly.
If you perform the linear regression with a programmed calculator,
additionally write down the correlation coefficient of the linear regression.

\item
On the straight line choose two points (A) and (B) 
and write down their coordinates:
energies $E_A$ and $E_B$ and frequencies $\nu_A$ and $\nu_B$.
Calculate the differences $\Delta E\equiv E_B-E_A$ and 
$\Delta\nu\equiv\nu_B-\nu_A.$

\item Calculate the ratio $h_\mathrm{LED}\equiv \Delta E/\Delta\nu$.
This is your evaluation of the Planck constant.
At correct solution of the problem you will obtain at least the correct order.

\item Calculate the ratio $X\equiv h_\mathrm{LED}/h$ 
between the measured by you with the present set-up value and 
the well-known value $h=6.62\times10^{-34}\,$J/Hz.
Calculate also $\epsilon=100\,(X-1)$, which is roughly speaking the error in percents.
For some purposes is more convenient to use frequency $\omega=2\pi\nu$.
In this case $\hbar\equiv h/2\pi=1.054\times10^{-34}\,$Js
is introduced and $h\nu=\hbar\omega.$

%%%%%%%%%%%%%%%%%%%%%%%%%%%
\section{Tasks L. Current voltage characteristics of LED}

\item 
Choose one of your multimeters to work as an ammeter,
and the other to work as a voltmeter.
Then measure the internal resistance of the ammeter $R_\mathrm{A}$
at its smallest current range.

\item
To study the current voltage characteristics $I(U)$ or I-V curves 
of the LEDs, connect the circuit from Fig.~\ref{Fig:U(I)}.
%%%%%%%%%%%%%%
\begin{figure}[ht]
\includegraphics[scale=0.5]{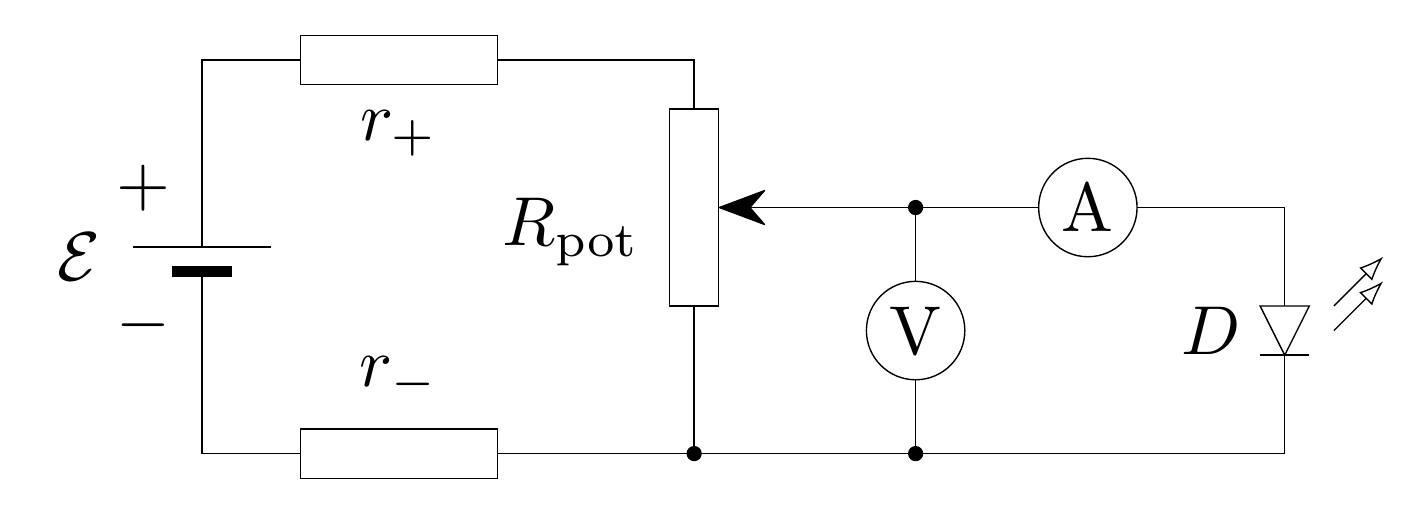}
\caption{
Circuit for study current-voltage characteristics.
Adjustable voltage regulator with potentiometer for voltage supplying 
to the LEDs and for investigation of $I(U).$
The voltage from the battery holder is applied to the potentiometer.
The LED is connected to the middle terminal and to one of the other terminals of the potentiometer.
%The LED is connected through a resistor $R$ to the middle terminal of the potentiometer and to one of the other terminals.
%The big value of the resistor $R$ allows to investigate this part of the V-I characteristic at which the LED starts or stops to emit light.
The ammeter is connected in series with the LED and the voltmeter is connected in parallel.
%If the LED does not light in one of the end positions of the potentiometer then you can change the polarity of the batteries.
%There are two resistors hidden in an insulated tubes before the crocodile clips of the battery holder leads.
}
\label{Fig:U(I)} 
\end{figure}
%%%%%%%%%%%%%

\item
We recommend to start with a red LED and sequentially study the others from the set.
The ammeter measures the current $I_\mathrm{A}$ through the LED $I$,
i.e. $I=I_\mathrm{A}$. 
The voltmeter shows voltage $U_\mathrm{V}$
and the voltage at the LED is $U=U_\mathrm{V}-R_\mathrm{A}I$.

\item
In this task we focus at small currents in the range $I\in (0.1,\, 10)\,\mu$A.
Rotating the axis of the potentiometer you have to look at the ammeter.
For each LED you have to record less than 10 current values below and above 
$\mathrm{1\,\mu A}$.
Observe the voltage values also, if necessary adjust the multi-meter range.
%For red LED voltmeter can be switched at 2000~mV (DCV),
%but for other one 20~V range have to be used.
Arrange your data in a table similar to Table~\ref{tbl:VAC}
%%%%%%%%
\begin{table}[h]
\begin{tabular}{c c }
		\tableline \tableline
		&  \\ [-1em]
		 $U_\mathrm{V}$~[V] & \hspace{5pt} $I$~[$\mu$A] 
		 \\ \tableline 
			&  \\ [-1em] 
			\dots	&	\dots \\
			\dots	&	\dots \\
\tableline \tableline
\end{tabular}
	\caption{Tabulated current voltage characteristics per LED,
	for each studied LED you need a separate table.}
	\label{tbl:VAC}
\end{table}

\item
Draw the I-V curve on a millimeter or squared paper.
Mark in a vertical arrow $U^*$ at the abscissa $U[\mathrm{V}]$.
Use the approximation of ideal ammeter $R_\mathrm{A}\approx 0$ 
when $U\approx U_\mathrm{V}.$

\item
Try to invent a method for determination of $U^*$
analyzing only the I-V curves at small currents 
and illustrate it graphically.

%%%%%%%%%%%%%%%%%%%
\section{Tasks XL. University students}
%%%%%%%%%%%%%%%%%%%

\item
\label{c-linear}
Perform the linear regression of the experimental points in the plot
$(\nu,\,q_eU^*)$ with a computer (if available)
and calculate more precisely $h_\mathrm{LED}$ and the correlation coefficient.
It is allowed to omit an experimental point with explanation of the reasons.

\item
If a point has been omitted, compare $h_\mathrm{LED}$
with the one obtained with all points.
Does the omission of a point decreases $\varepsilon$ and increases the correlation coefficient?

\item
Perform the order estimation of the errors in measurement of $U^*$.
You have at least 5 measurements.

\item Evaluate the error in the measurement of the diffraction angle $\theta$
and relate to it the wavelength $\lambda$.

\item
Evaluate $A$ in eV both from \ref{c-linear} (if available) and from \ref{linear}.

\item Consider what gives larger uncertainty for the final result
inaccuracies in $\Delta U^*$ and $\Delta\theta$
or the approximation that the band gap parameter 
$-A$ is the same constant for all LEDs.

\item 
The use of human eye inserts subjective uncertainties.
Invent a numerical method to determine the voltage $U^*$
analyzing $I$-$V$ curves (current-voltage characteristics).
Apply polynomial regression in some interval
and fit to exponent in another.
For example, use quadratic polynomial (parabolic) fit
\be I=a U^2+b U+c= a(U-\tilde{U})^2+(c-b^2/4a) \ee
for currents in the interval $I\in (1,\, 10)\,\mu$A
and calculate the position of the minimum $\tilde U=-b/2a$.
Calculate $|\tilde U-U^*|/U^*$ for the LEDs you study.

\item
Compare the obtained $\tilde{U}$ from the previous task and $U^*$ from \ref{item:U*} for all LEDs.
Present the errors between the two types of measurements in percent.

%%%%%%%%%%%%%%%%%%%%%%%%%%%%%%%%%%%%%%%
\section{Problems for Further Work After the Olympiad}

\item
The price of LEDs is very low. 
You can order LEDs with marked by the manufacturer wavelength and compare 
with your determination of the wavelength.
We expect 5-10\% accuracy.
If you find a good LED, please let us know at \url{epo@bgphysics.eu}.

\item What accuracy of determination of the wavelength $\lambda$ can be reached if you use lasers?

\item
To Table~\ref{tbl:VAC} add 3 additional columns to the right:
the correction $R_\mathrm{A}I$,
the corrected LED voltage $U=U_\mathrm{V}-R_\mathrm{A}I$,
and the LED resistance $R(U)\equiv U/I$ in M$\Omega$ as is demonstrated
in Table~\ref{tbl:VAC_extended}.
%%%%%%%%
\begin{table}[h]
\begin{tabular}{c c c c c}
		\tableline \tableline
		&  \\ [-1em]
		 $U_\mathrm{V}$~[V] 
		 & \hspace{5pt} 
		 $I$~[$\mu$A] 
		  & \hspace{5pt} 
		  $R_\mathrm{A}I$~[V] 
		   & \hspace{5pt} 
		    $U$~[V] 
		    & \hspace{5pt} 
		    R~[M$\Omega$]
		 \\ \tableline 
			&  \\ [-1em] 
			\dots	&	\dots & \dots	&	\dots & \dots \\
			\dots	&	\dots & \dots	&	\dots & \dots \\
\tableline \tableline
\end{tabular}
	\caption{Complete current voltage characteristics in which finite internal resistance of the ammeter is taken into account.}
	\label{tbl:VAC_extended}
\end{table}
Compare the minimal resistance from the table $R_\mathrm{min}$ 
with the resistance of the voltmeter
$R_\mathrm{V}$ in the used range.

\item If the current through the LED $I$ is large enough its resistance $U/I$ is very small and much smaller than the voltmeter internal resistance.
In this case it is necessary to use the connection depicted at 
Fig.~\ref{Fig:I-V_high_currents}. 
%%%%%%%%%%%%%%
\begin{figure}[ht]
\includegraphics[scale=0.5]{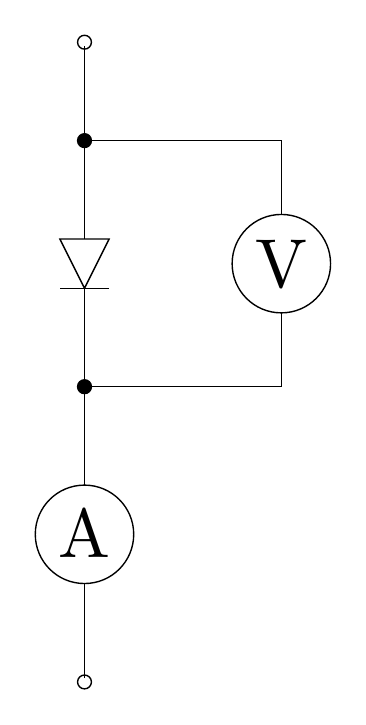}
\caption{Connection for high current and low LED resistance.
In this case $U=U_\mathrm{V}$
and the current through the LED
$I=I_\mathrm{A}-U/R_\mathrm{V}$.
When the voltmeter internal resistance is high enough 
$1/R_\mathrm{V}\approx 0$,
the corrections to the current $U/R_\mathrm{V}$
are negligible.
}
\label{Fig:I-V_high_currents} 
\end{figure}
%%%%%%%%%%%%%

\item If laser light falls on a vacuum lamp you can reach the accuracy 
of the first measurement of the Planck constant $h$ by Millikan.~\cite{Millikan}
Check what the accuracy of the determination of $h$ is in
the contemporary university labs.

\item Send to the address of the Olympiad epo@bgphysics.eu your responses, recommendations, criticism and suggestions, which you think would help the EPO9 organizers.
Negative impressions are also welcome and will be taken into account. 

\item Re-derive and program the solution equations for the parabolic fit 
of a set of experimental data $I=aU^2+bU+c$. 
Those are conditions for the minimization of the function
$$f(a,b,c)\equiv\sum_{i=1}^N (aU_i^2+bU_i+c-I_i)^2,$$
where summation is on the all $N$ points.
The conditions for minimum require the annulation of the first derivatives
$\partial_af=0$, $\partial_bf=0$, and $\partial_cf=0$, 
read 
$$\langle U^{2+m}\rangle a+\langle U^{1+m}\rangle b
+\langle U^{0+m}\rangle c= \langle U^{m}I\rangle,$$
where brackets denote averaging, for example 
$$\langle U^{m}I\rangle=\frac1{N}\sum_{i=1}^N  U_i^mI_i,
\qquad
\langle U^{k}\rangle=\frac1{N}\sum_{i=1}^N  U_i^k,
\quad k=0,1,2,3,4.
$$
In such a way we have a system of 3 linear equations, 
$m=0,\,1,\,2$, for the parameters
$a$, $b$ and $c$ which solutions can be easily programmed and of course 
it is implemented in many software programs addressed for users.

\item Draw the parabolic fit on the plot with experimental data $(U_i,\,I_i)$
and draw a small circle at the minimum of the parabola $\tilde U=-b/2a$.
Try to invent a better method for determination of the $U^*$
used for determination of Planck constant.

\item Most easily multimeters can be burnt when are used as ammeters.
That is why this task can be done after the Olympiad.
Every time start with the maximal fused current range.
Switch one of multimeters as ammeter and measure the maximal current
$I_\mathrm{max}$ which can give our voltage source.
Can we evaluate the internal resistance of the batteries 
$r_\mathrm{bat}$
using 
$I_\mathrm{max}=\mathcal{E}/(r+R_\mathrm{A}+3\,r_\mathrm{bat})$?

\end{enumerate}

%%%%%%%%%%%%%%%%%
\section{Epilogue}
EPO8 is held in a very difficult for the whole world conditions.
Approximately half of the participants are on-line.
The organizers of EPO8 would like to thank everyone 
who helped in the preparation of this wonderful competition especially
the president of the Society of Physicists of Macedonia, Prof.~Lambe Barandovski
for ensuring on-spot participation in Skopje.
We are waiting you at EPO9; next year same time.

%%%%%%%%%%%%%%%%%

%%%%%%

%\end{document}

\clearpage

\appendix

\section{Solutions to the EPO8 Problems by the Team from Toronto, Canada}

\includepdf[pages=-,width=\textwidth]{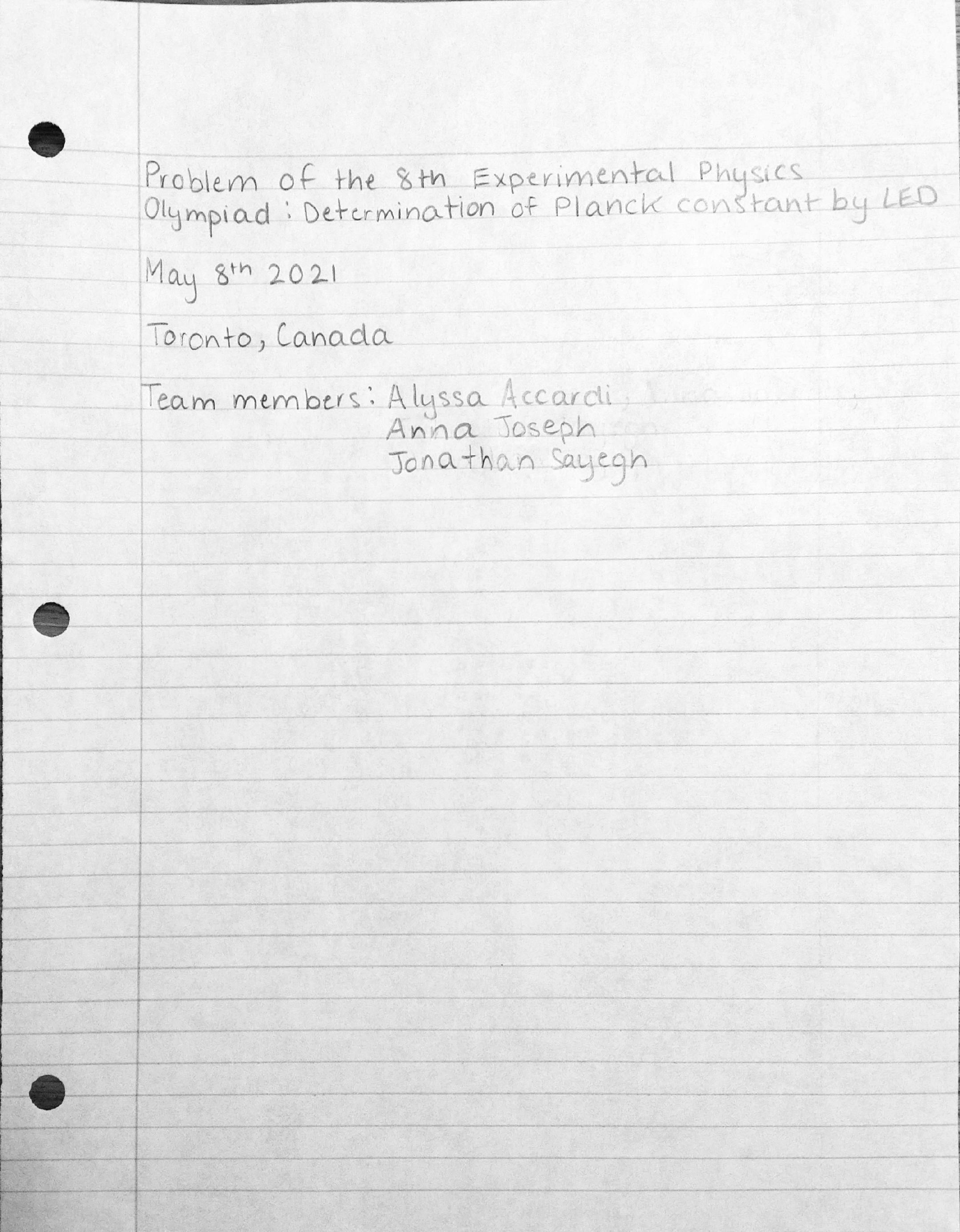}

\end{document}